\documentclass[twocolumn]{aastex6}
\RequirePackage{lineno}
\usepackage{amsmath}
\usepackage{amssymb}
\usepackage{amsthm}
\usepackage{natbib,aasdefs,url,bm}
\usepackage{array}
\usepackage{float}
\usepackage{graphicx}
\usepackage{subfigure}
\usepackage{color}

\newcommand{\pd}{\partial}  
\newcounter{ichi}
\newcounter{ni}
\newcounter{san}
\newcounter{yon}
\def\be{\begin{equation}}
\def\ee{\end{equation}}
\def\ba{\begin{eqnarray}}
\def\ea{\end{eqnarray}}

\slugcomment{}

\shorttitle{Late-Time High-Energy Signatures from Compact Remnants of Double Neutron Star Mergers}
\shortauthors{Murase et al.}

\begin{document}

\title{
Double Neutron Star Mergers and Short Gamma-Ray Bursts:\\
Long-Lasting High-Energy Signatures and Remnant Dichotomy
}
\author{Kohta Murase\altaffilmark{1,2,3,4}, Michael W. Toomey\altaffilmark{1}, Ke Fang\altaffilmark{5}, Foteini Oikonomou\altaffilmark{1,2,3,6}, Shigeo S. Kimura\altaffilmark{1,2,3},\\ 
Kenta Hotokezaka\altaffilmark{7}, Kazumi Kashiyama\altaffilmark{8}, Kunihito Ioka\altaffilmark{4}, Peter M\'esz\'aros\altaffilmark{1,2,3}
}
\altaffiltext{1}{Department of Physics, The Pennsylvania State University, University Park, PA 16802, USA}
\altaffiltext{2}{Department of Astronomy \& Astrophysics, The Pennsylvania State University, University Park, PA 16802, USA}
\altaffiltext{3}{Center for Particle and Gravitational Astrophysics, The Pennsylvania State University, University Park, PA 16802, USA}
\altaffiltext{4}{Center for Gravitational Physics, Yukawa Institute for Theoretical Physics, Kyoto University, Kyoto, Kyoto 606-8502, Japan}
\altaffiltext{5}{Department of Astronomy; Joint Space-Science Institute, University of Maryland, College Park, MD 20742, USA}
\altaffiltext{6}{European Southern Observatory, Karl-Schwarzschild-Str. 2, Garching bei M\"unchen D-85748, Germany}
\altaffiltext{7}{Department of Astrophysical Sciences, Princeton University, Princeton, NJ 08544, USA}
\altaffiltext{8}{Department of Physics, The University of Tokyo, Bunkyo, Tokyo 113-0033, Japan}


\begin{abstract}
The recent detection of gravitational waves and electromagnetic counterparts from the double neutron star merger event GW+EM170817, supports the standard paradigm of short gamma-ray bursts (SGRBs) and kilonovae/macronovae.
It is important to reveal the nature of the compact remnant left after the merger, either a black hole or neutron star, and their physical link to the origin of the long-lasting emission observed in SGRBs. The diversity of the merger remnants may also lead to different kinds of transients that can be detected in future. 
Here we study the high-energy emission from the long-lasting central engine left after the coalescence, under certain assumptions. In particular, we consider the X-ray emission from a remnant disk and the non-thermal nebular emission from disk-driven outflows or pulsar winds. We demonstrate that late-time X-ray and high-frequency radio emission can provide useful constraints on properties of the hidden compact remnants and their connections to long-lasting SGRB emission, and we discuss the detectability of nearby merger events through late-time observations at $\sim30-100$~d after the coalescence.
We also investigate the GeV-TeV gamma-ray emission that occurs in the presence of long-lasting central engines, and show the importance of external inverse-Compton radiation due to up-scattering of X-ray photons by relativistic electrons in the jet. 
We also search for high-energy gamma-rays from GW170817 in the {\it Fermi}-LAT data, and report upper limits on such long-lasting emission.
Finally, we consider the implications of GW+EM170817 and discuss the constraints placed by X-ray and high-frequency radio observations. 
\end{abstract}


\section{Introduction}
The gravitational wave (GW) event, GW170817, reported by the advanced Laser Interferometer Gravitational-Wave Observatory (LIGO) and the advanced Virgo Observatory, has been established as the first observation of a GW signal from a neutron star -- neutron star (NS-NS) merger~\citep{TheLIGOScientific:2017qsa}. 
Almost simultaneously with the GW signal, the Fermi Gamma-ray Burst Monitor ({\it Fermi}-GBM) and the Anticoincidence Shield for the Spectrometer for the International Gamma-Ray Astrophysics Laboratory (INTEGRAL) detected a short gamma-ray burst (SGRB), GRB 170817A, at a consistent location~\citep{GBM:2017lvd,Savchenko:2017ffs}. An ultraviolet, optical, and infrared (IR) counterpart of the burst were also detected, hours later, allowing for accurate identification of the host galaxy, NGC 4993, at a distance of $\sim40$~Mpc~\citep[see][and references therein]{Monitor:2017mdv}.  

The optical/IR counterpart of GW170817 is consistent with energy injection from the radioactive decay of neutron-rich heavy nuclei with mass $M\sim0.02-0.05~M\odot$~\citep[e.g.,][]{Arcavi:2017xiz,Chornock:2017aa,Coulter:2017wya,Cowperthwaite:2017dyu,Drout:2017aa,Kasliwal:2017aa,Kilpatrick:2017mhz,McCully:2017lgx,Nicholl:2017ahq,Pian:2017gtc,2017arXiv171005445R,Shappee:2017zly,Smartt:2017fuw,Tanaka:2017aa,Tanvir:2017aa,Utsumi:2017aa}, and the observational data are well explained by kilonova/macronova emission that was theoretically predicted~\citep{LP98a,Kulkarni:2005jw,MMD10a} and had been anticipated as a GW counterpart~\citep{Sylvestre:2003vc}. This observation also supports the simple scenario that $r$-process elements observed on the Earth dominantly originate from compact mergers involving neutron stars~\citep{Lattimer+76tproc,ELP89a,Wanajo:2014wha}. See \cite{Metzger:2017wot} for a summary of the kilonova developments. 
 
NS-NS mergers have also been believed to be the progenitors of SGRBs~\citep{1986ApJ...308L..43P,ELP89a,Meszaros+92tidal,NPP92a}. The precise origin of the gamma-rays in GRB 170817A is however under debate. If a canonical SGRB jet had been directed to the Earth (i.e. on-axis GRB events), that event would have produced exceedingly bright gamma-rays and multi-wavelength afterglow emission~\citep[e.g.,][]{GBM:2017lvd,Savchenko:2017ffs,Fong:2017aa,Kasliwal:2017aa}. The fact that the observed GRB prompt emission was weak has led to intense discussions about several scenarios such as off-axis emission from a canonical SGRB jet, or on-axis/off-axis emission from a structured jet with a wide-opening angle mildly relativistic component, and breakout emission from a mildly relativistic cocoon resulting from the jet-ejecta interaction~\citep[e.g.,][]{Bromberg:2017crh,Burgess:2017aa,Gottlieb:2017pju,Ioka:2017nzl,Kasliwal:2017aa,Lamb:2017ybq,2017ApJ...848L..34M,2018MNRAS.474L...7S,Shoemaker:2017nqv,Troja:2017nqp,2017arXiv171000275X,Zhang:2017lpb}. 
On the other hand, the late-time X-ray and radio data can be most naturally explained by off-axis afterglow emission from the canonical SGRB jet~\citep[e.g.,][]{Alexander:2017aly,Evans:2017aa,Hallinan:2017aa,Margutti:2017aa,Troja:2017nqp}, supporting the connection between SGRBs and NS-NS mergers. 

Whether the merger remnant is in general a black hole or neutron star is an open question. 
Although numerical general relativity simulations have suggested that the birth of a rapidly-rotating, massive NS just after the coalescence is ubiquitous~\citep[e.g.,][]{Shibata:1999wm,Shibata:2005ss,Sekiguchi:2011zd,Hotokezaka:2013iia,Kiuchi:2017zzg}, the fate of the remnant depends on some details such as the equation of state (EoS) and the total mass of the two NSs. 
Later, the massive NS may collapse into a black hole (BH), and the BH may be accompanied by a remnant disk. Alternatively, the massive NS may be long-lived and may power the merger ejecta via pulsar winds that become Poynting-dominated at late times, which could be the case especially for low-mass NS-NS binaries. 
See \cite{Shibata:2017xdx,Margalit:2017dij} for discussions about GW170817. 

Long-lasting activities of such a merger remnant may be imprinted in the observed high-energy emission. Indeed, some indirect clues have been found in the SGRB data. Whereas SGRBs are mainly detected as short hard bursts with a duration of $\lesssim2$~s~\citep[e.g.,][]{2004RvMP...76.1143P,Nak07a,2009ARA&A..47..567G,Ber14a,Ackermann:2013zfa}, it is known that a good fraction of SGRBs have late-time emission such as extended softer emission, X-ray flares, and plateaus~\citep{EBP09a,CMM10a,MCG11a,Swift11a,ROM13a,KYS15a,KBG15a,SwiftGRB16a,KIS17a}, and even longer mysterious X-ray excess emission was reported for GRB 130603B~\citep{Fong2014}, which is known to be coincident with the kilonova/macronova candidate~\citep[][]{Berger:2013wna,Tanvir:2013pia}. Resulting GeV-TeV gamma-ray emission~\citep{Wang:2006eq,MTY11a} and high-energy neutrino emission~\citep{MN06a,KMM17b} have also been studied.
These components can be most naturally interpreted as signatures of the long-lasting central engine.
The engine can be a spinning BH with a fall-back disk~\citep[e.g.,][]{Rosswog:2006rh,Nakamura:2013hda,KI15a}, where the jet may be powered by the Blandford-Znajek mechanism~\citep{BZ77a}. Alternatively, the engine could be a long-lived spinning NS or massive pulsar/magnetar if the EoS allows a relatively large maximum non-rotating NS mass~\citep[e.g.,][]{DL98a,ZM01a,Dai:2006aa,Fan:2006zx,MQT08a}. In the latter model, even brighter transients could be produced by the pulsar/magnetar wind nebula hours to days after the merger~\citep{Yu:2013kra}.
However, the physical link between long-lasting SGRB emission and compact remnants of NS-NS mergers remains missing. High-energy emission can be useful to reveal the compact remnants and their physical roles.    

The organization of this work is as follows. In Section~2, we consider the high-energy implications of compact merger remnants at late times. In particular, we study high-energy emission from a BH with a remnant disk and a long-lived pulsar, and investigate X-ray, radio, and gamma-ray signatures. Then, we discuss X-ray and radio implications of GW+EM170817. In Section~3, we consider external inverse-Compton radiation by a relativistic jet as one of the gamma-ray signals detectable by {\it Fermi} and imaging atmospheric Cherenkov telescopes. We also search for GeV-TeV gamma-ray counterparts of GW170817 within the {\it Fermi}-LAT data, and discuss the  implications of the results of this search. In Section~4, we summarize our results and describe prospects for future multi-messenger observations.     

\section{Hidden Compact Remnants and Long-Lasting Energy Injections} 
As a result of NS-NS mergers, the mass ejection happens through tidal stripping from the NSs, shock heating at the interface of the merging NSs, and disk-driven wind/outflow mass losses mainly via viscous heating and magnetohydrodynamic turbulence~\citep[see reviews, e.g.,][and references therein]{Rosswog:2012fn,Fernandez:2015use,Tanaka:2016sbx}. Observations of GW+EM170817 consistently suggest a post-merger ejecta mass of $M\sim0.02-0.05~M_{\odot}$. The ejecta are heated by radioactive nuclei synthesized at the violent coalescence, leading to an optical-IR emission, the so-called kilonovae/macronovae~\citep{Fernandez:2015use,Tanaka:2016sbx,Metzger:2017wot}. Throughout this work, we use the values of the post-merger ejecta mass and velocity, motivated by the observations of GW+EM170817. However, our purpose is rather to provide a general study on possible roles of the central engine, so we consider a variety of high-energy signatures, including cases that are very different from the case of GW+EM170817.  

In addition to radioactive decay heating, one expects additional energy injection by a BH with a disk or long-lived NS. For the purpose of revealing associated non-thermal signatures, we consider the following simplified toy model. 
The evolution of the thermal energy (${\mathcal E}_{\rm th}$) is described by
\begin{equation}
\frac{d{\mathcal E}_{\rm th}}{dt}=Q-\frac{{\mathcal E}_{\rm th}}{R}\frac{dR}{dt}-L_{\rm th},
\end{equation}
where $R$ is the ejecta radius and $V=dR/dt$ is the ejecta velocity.  
To calculate the evolution of quantities such as ${\mathcal E}_{\rm th}$, $V$ and $R$, we follow \cite{KMB16a} and \cite{Omand:2017hgg}, which were calibrated to calculate supernova light curves~\citep{1982ApJ...253..785A}.
Here $Q=Q_{\rm rah}+Q_{\rm nth}$ is the total thermal energy deposition rate, and $Q_{\rm nth}=f_{\rm nth}^{\rm dep}\dot{\mathcal E}_{\rm nth}$ is the heating rate by the absorption and scattering of non-thermal radiation (where $f_{\rm nth}^{\rm dep}$ is the energy deposition fraction for thermal energy, and $\dot{\mathcal E}_{\rm nth}$ is the energy production rate).
For the radioactive heating rate, $Q_{\rm rah}$, we adopt the parametrization of \cite{Hotokezaka:2017dbk}. The total energy production rate including gamma-rays, $\alpha$ particles, and spontaneous fission may be higher than this parametrization by a factor of $3-10$~\citep[cf.][]{MMD10a,Wanajo:2014wha}, but the resulting luminosity depends on microphysical details of the thermalization~\citep{Barnes:2016umi}. For the purpose of this work, details of the treatment are not important, and the approximate modeling of the optical/IR emission is sufficient within uncertainty of nuclear physics. We adopt the normalization, $q_{\rm rah}={10}^{10}~{\rm erg}~{\rm cm}^{-3}~{\rm g}$ at $\bar{Z}=100$ and $\bar{A}=200$, following \cite{Wanajo:2014wha}, and the late-time bolometric light curve of GW+EM170817 is basically reproduced with a post-merger ejecta mass of $M\sim0.02~M_\odot$ with an ejecta velocity of $V\sim0.2~c$, which is consistent with other studies on GW+EM170817. 

The luminosity of thermal radiation is calculated from
\begin{equation}
L_{\rm th}=\frac{{\mathcal E}_{\rm th}}{t_{\rm esc}^{\rm ej}}
\end{equation}
where $t_{\rm esc}^{\rm ej}={\rm min}[1,\tau_T^{\rm ej}]R/c$ is the escape time of the thermal radiation, $\tau_{\rm T}^{\rm ej}\approx{MK}/{4\pi R^2}$ is the optical depth to scattering with electrons in atoms (where the coefficient depends on details of the velocity profile of the ejecta), and $K$ is the opacity at optical and IR bands, which depends on the mass composition and ionization state of the post-merger ejecta. 
The observations of GW+EM170817 suggest that the optical/IR emission consists of blue and red components. Although details of their physical origin are still under debate, one of the interpretations is that the latter comes from $r$-process nuclei entrained in the dynamical ejecta. In such neutron-rich, dynamical ejecta, a high opacity of $K\sim5-10~{\rm cm}^2~{\rm g}^{-1}$ is expected~\citep[e.g.,][]{KBB13a,Tanaka:2017aa}, whereas relatively lanthanoid-free disk winds may have a lower opacity of $K\lesssim1~{\rm cm}^2{\rm g}^{-1}$~\citep[e.g.,][]{BK13a}. 

To study high-energy signatures produced by a compact remnant, we consider two examples: a BH with a fall-back disk and a long-lived pulsar. 
In the former scenario, the accretion disk around a stellar-mass BH is naturally accompanied by X-rays as seen in ultra-luminous X-ray sources. In addition, as often considered in the context of GRBs, dissipation via shocks or magnetic reconnection may naturally occur in ultrafast disk-driven outflows or in relativistic pulsar winds. Then particle acceleration is expected to occur, and we calculate the resulting non-thermal radiation, including effects of electromagnetic cascades. We solve the following kinetic equations~\citep{MKK15a}:  
\begin{eqnarray}\label{eq:cascade}
\frac{\pd n_{E_e}^e}{\pd t} &=& \frac{\pd n_{E_e}^{(\gamma\gamma)}}{\pd t} 
- \frac{\pd}{\pd E_e} [(P_{\rm IC}+P_{\rm syn}+P_{\rm ad}) n_{E_e}^e] + \dot{n}_{E_e}^{\rm inj},\nonumber\\
\frac{\pd n_{E_\gamma}^\gamma}{\pd t} &=& -\frac{n_{E_\gamma}^{\gamma}}{t_{\gamma \gamma}} - \frac{n_{E_\gamma}^{\gamma}}{t_{\rm esc}}
+ \frac{\pd n_{E_\gamma}^{(\rm IC)}}{\pd t} 
+ \frac{\pd n_{E_\gamma}^{(\rm syn)}}{\pd t} +\dot{n}_{E_\gamma}^{\rm inj},
\end{eqnarray}
where $t_{\gamma\gamma}$ is the two-photon annihilation time, $t_{\rm esc}$ is the photon escape time from the nebular region, $\pd n_{E_e}^{(\gamma\gamma)}/\pd t$ is the electron-positron injection rate via $\gamma\gamma\rightarrow e^+e^-$, $P_{\rm IC}$ is the inverse-Compton energy-loss rate, $P_{\rm syn}$ is the synchrotron energy-loss rate, and $P_{\rm ad}$ is the adiabatic energy-loss rate.
Particle injection rates, $\dot{n}_{E_\gamma}^{\rm inj}$ and $\dot{n}_{E_e}^{\rm inj}$, are determined by energy injection from the central engine. 
We here include relic particles as well as freshly injected particles, which is relevant when the particle injection rate declines more rapidly than $\propto t^{-1}$.  

High-energy emission from compact remnants of NS-NS mergers is attenuated during the propagation of photons in the ejecta, and we approximately implement an energy-dependent opacity. Note that X-rays, MeV gamma-rays, GeV-TeV gamma-rays, and radio waves are attenuated via {\it different} physical processes. Thus, multi-wavelength and multi-messenger observations from radio waves to GeV-TeV gamma-rays can provide us with independent probes of the merger ejecta. 

For X-rays, we exploit the mass energy-transfer coefficient, $\hat{K}_X$, where we use the mass energy-transfer coefficient data provided by NIST~\footnote{\url{https://www.nist.gov/pml/x-ray-mass-attenuation-coefficients}}. 
For simplicity, we consider two cases. In the high-opacity case ($K=10~{\rm cm}^2~{\rm g}^{-1}$ in the optical/IR band), which is expected in the neutron-rich ejecta with $Y_e\sim0.1-0.2$, we assume the mass composition averaged over xenon and gold. In the modest opacity case ($K=0.3~{\rm cm}^2~{\rm g}^{-1}$ in the optical/IR band), which is expected in the lanthanoid-free ejecta with $Y_e\sim0.3-0.4$, we assume the mass composition averaged over selenium and iron. The difference becomes irrelevant once the ejecta are optically thin for X-rays. Note that the effective optical depth for X-rays is given by $f_X=\hat{K}_X\rho R$, and the interaction with matter can essentially be treated as absorption. With $f_X=1$, the hard X-ray breakout time is analytically estimated to be
\begin{eqnarray}
t_{HX-{\rm thin}}&\simeq&4.2\times{10}^6~{\rm s}~{\left(\frac{\hat{K}_X}{100~{\rm cm}^2~{\rm g}^{-1}}\right)}^{1/2}\nonumber\\&\times&{\left(\frac{M}{0.02~M_\odot}\right)}^{1/2}{\left(\frac{V}{0.2~c}\right)}^{-1},
\end{eqnarray}
where $\hat{K}_X\sim100~{\rm cm}^2~{\rm g}^{-1}$ is a typical value at a photon energy of $E\sim10$~keV but higher at lower energies, and $\rho\approx[(3-\delta)/(4\pi)](M/R^3)$ is used with an inner density profile of $\delta\sim1$. We mainly consider cases where the post-merger ejecta are not ionized (especially for electrons in the inner shells of constituting atoms) by high-energy emission from the disk emission or wind nebular emission, in which our treatment gives conservative estimates of the X-ray flux.  

Gamma-rays have an advantage in that the attenuation cross section is much smaller than that of the bound-free absorption in the keV range. Also, the optical depth for high-energy gamma-rays does not depend on details of the ionization state. Compton scattering is dominant in the MeV range, and the MeV gamma-rays escape at
\begin{eqnarray}
t_{\rm MeV\gamma-{\rm thin}}&\simeq&7.3\times{10}^4~{\rm s}~{\left(\frac{\hat{K}_{\rm Comp}}{0.03~{\rm cm}^2~{\rm g}^{-1}}\right)}^{1/2}\nonumber\\&\times&{\left(\frac{M}{0.02~M_\odot}\right)}^{1/2}{\left(\frac{V}{0.2~c}\right)}^{-1},
\end{eqnarray}
in the large inelasticity limit, which is valid for $E\gtrsim1$~MeV (where the correction for bound states is also small). Bethe-Heitler pair-production becomes important above the pair-production threshold, which is more important for heavier nuclei. The effective optical depth is given by $f_{\rm BH}\approx(\tilde{Z}_{\rm eff}/2+\mu_e^{-1})\kappa_{\rm BH}\sigma_{\rm BH}^{(p)}(\rho/m_H) R$, where $\tilde{Z}_{\rm eff}=\Sigma_i (2Z_i^2/A_i)x_i$, $\kappa_{\rm BH}\sim1$ is the inelasticity, and $\sigma_{\rm BH}^{(p)}\sim{10}^{-26}~{\rm cm}^2$ is the Bethe-Heitler cross section for $\gamma p\rightarrow pe^+e^-$ at GeV energies. The condition $f_{\rm BH}=1$ gives the GeV gamma-ray breakout time, which is
\begin{eqnarray}
t_{\rm GeV\gamma-\rm thin}&\simeq&1.6\times{10}^5~{\rm s}~{([\tilde{Z}_{\rm eff}+2\mu_e^{-1}]/50)}^{1/2}\nonumber\\
&\times&{(M/0.02~M_\odot)}^{1/2}{(V/0.2~c)}^{-1}.
\label{gammatime}
\end{eqnarray}

We calculate not only X-rays and gamma-rays, but also radio waves, following~\cite{Murase:2016sqo}. We take into account synchrotron self-absorption and other relevant absorption processes. In particular, the free-free absorption process is important. We expect that the merger ejecta are nearly neutral at late times. To get a relatively conservative estimate of the resulting radio flux, for the merger ejecta, we assume a singly ionized material with a black body temperature, ignoring the charge-screening effect. Radio emission could more easily escape if the ejecta material is more neutral or the temperature is higher than that of a black body, or the composition is lighter.

\subsection{Black hole with a remnant disk}
SGRBs are known to show long-lasting activity. X-ray extended emission continues for $\sim100-300$~s~\citep[e.g.,][]{Swift11a,KYS15a,KBG15a}, and X-ray flares have been observed even $\sim{10}^4-{10}^5$~s after the prompt emission~\citep[e.g.,][]{CMM10a,MCG11a}, and plateau emission continues for $\sim{10}^4$~s~\citep[e.g.,][]{EBP09a,ROM13a,KIS17a}. Even longer mysterious X-ray excess emission, which cannot be explained by the simple energy-injection afterglow model, was reported for GRB 130603B~\citep{Fong2014}. 
The origin of the long-lasting, flaring or plateau emission is thought to be internal. It is sometimes called late prompt emission, and is attributed to internal dissipation in SGRB jets~\citep[e.g.,][]{GGN07a,MTY11a}. The most common explanation has been that it originates from the long-lasting central engine~\citep[e.g.,][]{Ioka:2004gy}. 

It is natural to expect that a fraction of the dynamical ejecta fall back onto a central compact object, and the fall-back material may power late-prompt emission of SGRBs~\citep{KI15a,KIN15b}. The mass fall-back rate can be written as
\begin{equation}
\dot{M}_d(t)=\frac{2M_{d}}{3t_{\rm eje}}{\left(\frac{t}{t_{\rm eje}}\right)}^{-5/3},
\end{equation}
where $M_{d}$ is the fall-back disk mass, and $t_{\rm eje}$ is the characteristic mass ejection time scale. The fall-back mass rate has been estimated to be $\dot{M}_0\sim{10}^{-4}-{10}^{-3}~M_\odot~{\rm s}^{-1}$ for $M_{d}\sim M \sim 0.03~M_\odot$ and $t_{\rm eje}\sim0.01-0.1$~s~\citep[e.g.,][]{Rosswog:2006rh,Rossi:2008eq,KI15a}. The post-merger ejecta mass may be larger than the dynamical ejecta mass, leading to a larger remnant mass, while a disk-driven outflow or its radiation may reduce or modify the fall-back accretion rate especially at late times~\citep{Fernandez:2014bra}. 

On the other hand, a large mass-accretion rate is necessary to explain SGRB long-lasting emission in such a BH disk model. X-ray plateaus and flares are observed at $t\sim{10}^4-10^5$~s, and the typical isotropic-equivalent luminosity of plateau emission is $L\sim{10}^{47}~{\rm erg}~{\rm s}^{-1}$. Assuming radiation efficiency $\epsilon_{\gamma}$, the jet efficiency, $\eta_j$, and the beaming factor, $f_b\approx\theta_j^2/2$ (where $\theta_j$ is the jet opening angle), we have $L=\epsilon_\gamma\eta_j \dot{M}_dc^2/f_b$, leading to 
\begin{equation}
\dot{M}_d\simeq5.6\times{10}^{-9}~M_\odot~{\rm s}^{-1}~f_{b,-2}\epsilon_{\gamma,-1}^{-1}\eta_{j}^{-1}L_{47}.
\end{equation} 
The accretion mass needed to explain plateau emission with duration $10^4-10^5$~s can be as small as $M_d\sim6\times({10}^{-5}-{10}^{-4})~M_\odot$. 
With a fall-back temporal index, the above mass accretion rate at $t=1$~s is extrapolated to be $\dot{M}_0\sim0.03~M_\odot~{\rm s}^{-1}$. Even with $\epsilon_\gamma\sim1$ and $\eta_j\sim1$ (which may be possible in the magnetically-dominated state of the disk), one needs $\dot{M}_0\sim3\times{10}^{-3}~M_\odot~{\rm s}^{-1}$. Thus, we would need a large accretion rate of $\dot{M}_0\gtrsim{10}^{-3}~M_\odot~{\rm s}^{-1}$ to explain the SGRB plateau emission. 

In this work, we parametrize the mass accretion rate to the remnant BH as 
\begin{equation}
\dot{M}_d=\dot{M}_0t^{-\alpha_{\rm acc}},
\end{equation}
and we adopt $\alpha_{\rm acc}=5/3$. To illustrate the results, we consider $\dot{M}_0={10}^{-3}~M_{\odot}~{\rm s}^{-1}$ as a baseline case, and take $\dot{M}_0={10}^{-2}~M_{\odot}~{\rm s}^{-1}$ as an optimistic case. It would be sufficient for us to demonstrate that late-time high-energy observations can provide constraints on $\dot{M}_d$, which is useful for the purpose of testing the origin of long-lasting SGRB emission and its connection to NS-NS mergers. 
The mass accretion rate is likely to largely exceed the Eddington rate at early times of the fall-back mass accretion. We here consider two phenomenological models, although details depend on the physics of super-Eddington accretion onto the remnant BH and dedicated radiative magnetohydrodynamical simulations will be necessary.

First, we consider the {\it disk emission} model, assuming that X-rays are produced by a corona of a fall-back disk or a disk itself. Analogously to the slim disk model for super-Eddington accretion flows (where photon trapping is effective), one may expect the disk luminosity to be 
\begin{eqnarray}
L_{\rm disk}^{\rm max}&=&\eta_{\rm rad} L_{\rm Edd}=\eta_{\rm rad}\frac{4\pi GM_{\rm rem} \mu_e m_Hc}{\sigma_T}\nonumber\\
&\simeq&1.0\times{10}^{40}~{\rm erg}~{\rm s}^{-1}~\left(\frac{\eta_{\rm rad}}{15}\right)\left(\frac{\mu_e}{2}\right)\left(\frac{M_{\rm rem}}{2.8~M_\odot}\right),\,\,\,\,\,\,\,\,\,\,
\label{Ldiskmax}
\end{eqnarray}
where $\eta_{\rm rad}\sim10-30$ is possible in the slim disk model with $\dot{M}_d\gg\dot{M}_{\rm Edd}$, depending on the ratio of the disk outer radius and inner radius~\citep{KFM08a,OMN05a}. In this work, we adopt $L_{\rm disk}^{\rm max}={10}^{40}~{\rm erg}~{\rm s}^{-1}$, with a spectrum $dL_{E}/dE\propto E^{-2}\exp(-E/E_{\rm cut})$ with $E_{\rm cut}=100$~keV at $E\geq1$~keV, motivated by modelling of the observed coronal emission from ultra-luminous X-ray sources~\citep[e.g.,][]{Kitaki:2017rlk}.
In the late phase, the accretion mode enters the sub-Eddington regime with a typical efficiency of $\sim0.1$. Note that we fix the spectral shape for simplicity. The cutoff energy may be lower and can be in the $\sim1-10$~keV range. We use $L_{\rm disk}={\rm min}[L_{\rm disk}^{\rm max},0.1 \dot{M}_dc^2$]. 
The transition occurs at $t_{\rm tr}\simeq1.4\times{10}^{6}~{\rm s}~{(\dot{M}_0/10^{-3}~M_\odot~{\rm s}^{-1})}^{3/5}{(L_{\rm disk}^{\rm max}/10^{40}~{\rm erg}~{\rm s}^{-1})}^{-3/5}$.

\begin{figure}[tb]
\begin{center}
\includegraphics[width=\linewidth]{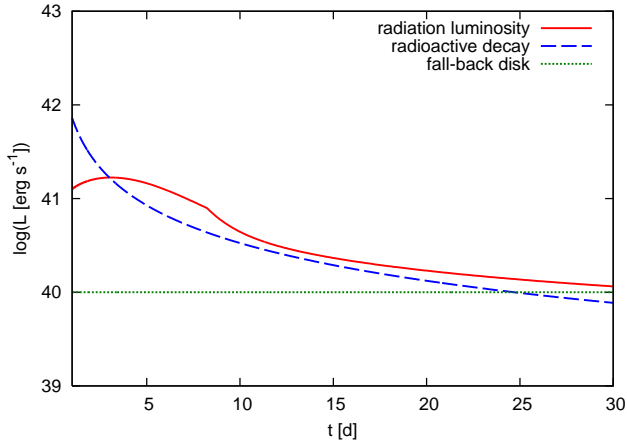}
\caption{Thermal bolometric luminosities from a NS-NS merger. We show the case where thermal radiation is mainly powered by radioactive decay of $r$-process elements and modified by X-ray emission from the disk with $L_{\rm disk}=10^{40}~{\rm erg}~{\rm s}^{-1}$ for demonstration purposes.
}
\label{BHlc}
\end{center}
\end{figure}

A sample of ``thermal'' bolometric light curves is shown in Figure~\ref{BHlc}, where the ejecta mass and velocity are assumed to be $M=0.02~M_\odot$ and $V=0.2~c$, respectively, and the thermal radiation essentially corresponds to kilonova/macronova emission with ${\mathcal T}\sim{10}^{3}-{10}^4$~K. 
In Figure~\ref{BHlc}, we show an example which illustrates that late-time bolometric light curves can in principle be modified by the disk emission. If a significant fraction of the accretion luminosity is converted into radiation even at early times, the thermal radiation should be modified, by which we can constrain the model. This case corresponds to $t_{\rm tr}\gtrsim30$~d, that is, $\dot{M}_0\gtrsim3\times{10}^{-3}~M_{\odot}~{\rm s}^{-1}~{(L_{\rm disk}^{\rm max}/10^{40}~{\rm erg}~{\rm s}^{-1})}$. 
Note that the disk emission does not affect the thermal bolometric light curve in our fiducial case with $\dot{M}_0={10}^{-3}~M_\odot~{\rm s}^{-1}$.

\begin{figure}[tb]
\begin{center}
\includegraphics[width=\linewidth]{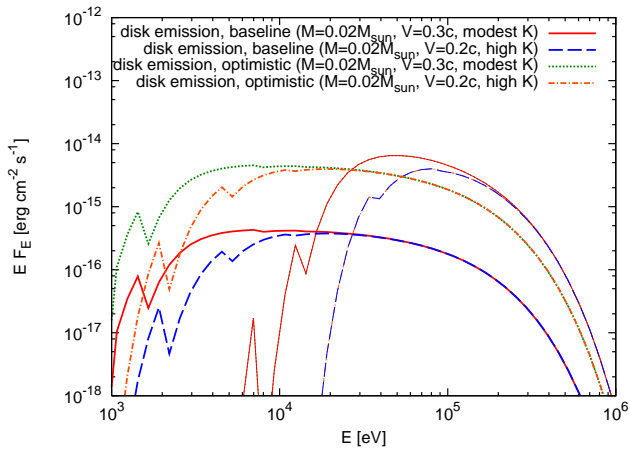}
\caption{X-ray spectra from a BH with a remnant disk in the disk emission model, at $t=10^6$~s (thin curves) and $t=10^7$~s (thick curves). The distance is set to $d=40$~Mpc. Note that the baseline and optimistic cases are degenerate at $t=10^6$~s.}
\label{BHspe}
\end{center}
\end{figure}

\begin{figure}[tb]
\begin{center}
\includegraphics[width=\linewidth]{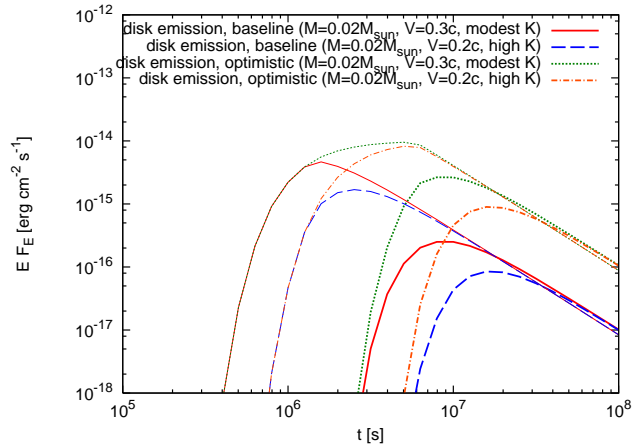}
\caption{X-ray light curves from a BH with a remnant disk in the disk emission model, for $E=3$~keV (thick curves) and $E=30$~keV (thin curves).}
\label{BHX}
\end{center}
\end{figure}

The results on X-ray spectra and light curves are shown in Figures~\ref{BHspe} and \ref{BHX}. 
X-ray spectra are significantly absorbed at early stages, which are unlikely to be observed until $t_{HX-\rm thin}\sim10-50$~d. We expect that the disk luminosity cannot exceed the Eddington luminosity by more than a factor of $\gtrsim10-100$, so detection of X-rays is possible only for nearby NS-NS mergers at late times. 
{\it Chandra}, with sensitivity $EF_E\sim{10}^{-15}~{\rm erg}~{\rm cm}^{-2}~{\rm s}^{-1}$, may enable us to observe the emission from a remnant BH with a disk left after a NS-NS merger up to $\sim30-50$~Mpc. 
The future mission, {\it Athena+}~\citep{Nandra:2013shg}, is planned to reach a limiting sensitivity of $EF_E\sim{10}^{-17}~{\rm erg}~{\rm cm}^{-2}~{\rm s}^{-1}$ in the $0.2-2$~keV range, by which the X-ray emission from the accretion onto the remnant BH is detectable up to $\sim300-500$~Mpc. 
Effects of the mass composition on the opacity can affect early-time fluxes by a factor of $\sim10$, which can be regarded as one of the uncertainties. 

For X-ray observations, deep measurements at $\sim t_{HX-\rm thin}$ are important. If the observational limits reach $L_{\rm disk}\lesssim L_{\rm Edd}$, we can examine whether the accretion mode is super-Eddington or not, which allows us to constrain $\dot{M}_d$ and test the remnant disk model as the central engine of long-lasting SGRB emission.  

\begin{figure}[tb]
\begin{center}
\includegraphics[width=\linewidth]{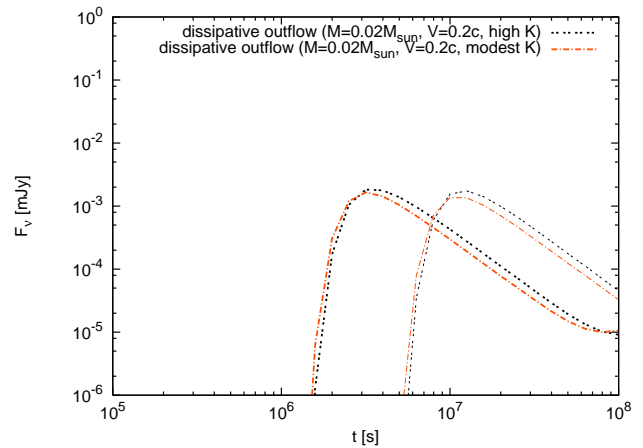}
\caption{High-frequency radio light curves from a BH with a remnant disk in the dissipative outflow model, for $\nu={10}^{10}$~Hz (thin curves) and $\nu=10^{11}$~Hz (thick curves).}
\label{BHradio}
\end{center}
\end{figure}

The above assumption that X-rays come from the disk or its corona may be reasonable when the mass accretion rate is not far from that of ultra-luminous X-ray sources. 
At earlier times, the disk radiation luminosity should not largely exceed the Eddington luminosity, but the outflow kinetic luminosity can be larger. A significant fraction of the disk mass, $\eta_w\sim0.1-1$, may be ejected back into space as an ultrafast outflow, perhaps by viscous heating~\citep[e.g.,][]{Dessart:2008zd,Fernandez:2013tya} and/or magnetohydrodynamic turbulence~\citep[e.g.,][]{Price&Rosswog06,Kiuchi+15} as well as radiation pressure~\citep[e.g.,][]{OMN05a,JSD14a,NSS17a}. 
The velocity of such ultrafast outflows is expected to be $V_{w}\sim0.05-0.3~c$, depending on mechanisms of the outflow production. 
The properties of disk-driven outflows at later times seem more uncertain. While in the presence of strong winds the accretion rate may decline more rapidly than the fall-back rate, higher mass-loss rates seem necessary to explain the long-lasting SGRB emission. 

In this work, as the second phenomenological model, we consider the {\it dissipative outflow} model for late-time disk-driven winds. 
The disk-driven wind may simply carry kinetic energy equal to that of the post-merger ejecta, but it may cause interactions with the dynamical ejecta if the late-time outflow has a high velocity $V_w>V_{\rm ej}$. We consider such a BH wind nebula formed by the shock. 
At sufficiently late times, radiation can escape into the ejecta and contributes to heating of the ejecta, and the shock velocity is estimated to be ${(L_w/[2\pi\rho R_w^2])}^{1/3}$ in the ejecta comoving frame. Note that the nebula size, $R_w$, should be limited by the ejecta radius, $R$. A fraction of the kinetic energy may be used for particle acceleration, where one can expect synchrotron emission as considered in pulsar wind nebulae. 
The kinetic energy of the outflow is written as
\begin{equation}
L_{w}=\eta_w\dot{M}_dV_w^2, 
\end{equation} 
and we set $\eta_w=0.3$ and $V_w=0.3~c$. In this work, we only consider the acceleration of primary electrons with a simple power-law injection. For simplicity, the spectral index is set to $s=2.2$, the electron injection Lorentz factor is fixed to $\gamma_{e,i}=1$, and the energy fractions of non-thermal electrons is assumed to be $\epsilon_e=0.1$. We also assume that $\epsilon_B=0.003$ of the outflow kinetic energy $M_wV_w^2/2$ is stored as the magnetic field, and we use a wind mass of $M_w=0.02~M_\odot$. While this is sufficient as an illustration, the value is highly uncertain.  

We have found that detecting X-ray and gamma-ray emission from a hidden BH disk wind is difficult for our fiducial parameters, simply because the mass accretion rate quickly declines with time. However, as shown in Figures~\ref{BHradio}, high-frequency radio emission may be detectable. For example, ALMA has sensitivity $F_\nu\sim0.01$~mJy at $\nu=100$~GHz. For our parameters, the radio flux at the peak time (at $\sim30-100$~d) is $F_\nu\sim{\rm a~few}~{\mu\rm Jy}~{(d/40~{\rm Mpc})}^{-2}$, and then it decreases as $F_\nu\propto t^{-5/3}$ following $\dot{M}_d$. Note that the synchrotron nebular spectrum in this BH disk wind model is approximately given by $F_\nu\propto \nu^{-s/2}\propto\nu^{-1.1}$. 
Specific implications of GW+EM170817 are described in Section~2.3.

\subsection{Long-lived spinning neutron star}
The birth of a rapidly-rotating, massive NS just after the coalescence seems ubiquitous, as has been found in a series of fully relativistic numerical simulations~\citep[e.g.,][]{Shibata:1999wm,Shibata:2005ss,Sekiguchi:2011zd,Hotokezaka:2013iia,Kiuchi:2017zzg}. The lifetime of the massive NS depends on the nuclear EoS, and a very long-lived spinning NS can be formed for sufficiently stiff EoSs especially in a low-mass NS-NS merger. Due to the large angular momentum of the binary system, the immediate remnant is typically expected to have an extremely rapid rotation with a rotation period of $P_i\sim1$~ms~\citep[but see][for the slower rotation of the NS core]{Ciolfi+17,Hanauske:2016gia}. The amplification of magnetic fields also naturally occurs via magnetic field winding, shear instability, and magneto-rotational instability~\citep{Price&Rosswog06,Zrake&MacFadyen13,Kiuchi+15,Kiuchi:2017zzg}. Although the formation of ordered magnetic fields~\citep[e.g., via the dynamo mechanism;][]{1992ApJ...392L...9D,2015ApJ...809...39G} is still under debate, the remnant NS may acquire a strong large-scale magnetic field with $B_*\gtrsim{10}^{15}$~G. 

While the set of spin-down parameters, $P_i$ and $B_*$, is uncertain, the phenomenological consequences of such a long-lived pulsar or magnetar have largely been investigated in various contexts, which include SGRBs, isotropic optical and X-ray counterparts of GW sources, and fast radio bursts~\citep[e.g.,][]{DL98a,ZM01a,Dai:2006aa,Fan:2006zx,MQT08a,Shibata:2011fj,ROM13a,Totani:2013lia,Yu:2013kra,Metzger:2014ks,Gao:2015rua,DOrazio:2015jcb,Murase:2016ysq,Piro+17}. Assuming the formula of \cite{Gruzinov:2004jc} for an aligned rotator, the injection luminosity and the energy injection time are given by the spin-down luminosity and the spin-down time as
\begin{eqnarray}
L_{\rm inj}\approx L_{\rm sd}&\approx&\frac{4\pi^4B_*^2R_*^6{P}^{-4}}{c^3}\nonumber\\
&\simeq&1.4\times10^{48}~{\rm erg}~{\rm s}^{-1}~B_{*,16.5}^2P_{-2}^{-4}R_{*,6}^6
\end{eqnarray}
and
\begin{equation}
t_{\rm inj}\approx t_{\rm sd}\approx\frac{P_i^2{\mathcal I}c^3}{2\pi^2B_*^2R_*^6}\simeq130~{\rm s}~B_{*,16.5}^{-2}P_{i,-2}^{2}R_{*,6}^{-4},
\end{equation}
where $P_i$ is the initial rotation period, $B_{*}$ is the dipole magnetic field at the surface, and ${\mathcal I}$ is the moment of inertia. 
In particular, the long-lived pulsar model has been very popular to explain the extended emission of SGRBs, and the observations can be well explained with $B_*\sim{\rm a~few}\times{10}^{16}$~G and $P_i\sim10$~ms~\citep{Gompertz:2013aka,Murase:2016ysq}. 
While this spin period seems longer than the typical value of $P_i\sim1$~ms, such values seem necessary to explain very late emission such as plateaus in the long-lived pulsar model~\citep{Fong2014}. Thus, allowing for arbitrary spin-down parameters, we consider both the modest and high opacity cases, as in the previous subsection. However, recent numerical studies have indicated that a long-lived NS is accompanied by long-lasting neutrino emission, leading to a higher value of the electron fraction, $Y_e$~\citep[e.g.,][]{Lippuner:2017bfm}. If this is the case, the modest opacity would be more favorable for long-lived NS remnants.  

Studies of Galactic pulsar wind nebulae suggest that almost all the spin-down power is extracted in the form of a Poynting flux, and eventually released as non-thermal nebular emission rather than thermal radiation. The Poynting-dominated pulsar wind is accelerated to relativistic speed although its evolution beyond the light cylinder has been under debate. In this work, following~\cite{MKK15a} and \cite{Omand:2017hgg}, we make the ansatz that microphysical parameters of embryonic pulsar wind nebulae are the same as those of young pulsar wind nebulae, in which it is assumed that electrons and positrons are accelerated with a broken power law with $s_1=1.5$ and $s_2=2.5$ with a break Lorentz factor $\gamma_b=10^5$. The magnetic energy fraction is set to $\epsilon_B=0.003$ and the rest of the energy ($\epsilon_e=1-\epsilon_B$) is used for the acceleration of electrons and positrons. 

In the very early stages of the nebular evolution, non-thermal emission can be thermalized even inside the nebula~\citep{Metzger:2014ks,Fang:2017tla}.  
At late times, given that the pulsar wind is highly relativistic beyond the light cylinder, the pair density quickly drops, and subsequently the thermalization of non-thermal photons starts to occur instead in the merger ejecta. When intra-source electromagnetic cascades do not occur in the so-called saturated cascade regime, it is relevant to solve kinetic equations to discuss high-energy signatures that can escape from the nebula and ejecta~\citep{MKK15a}. 
  


\begin{figure}[tb]
\begin{center}
\includegraphics[width=\linewidth]{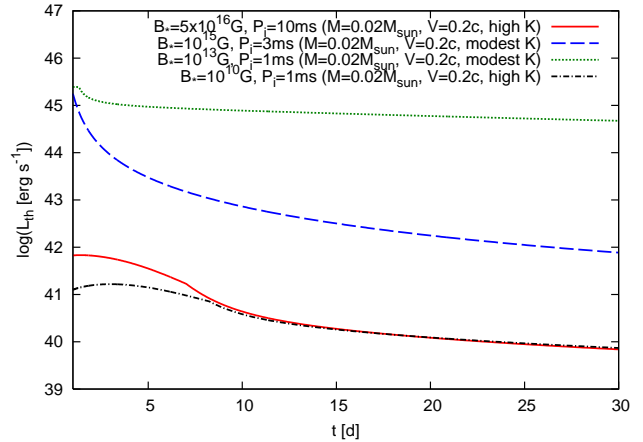}
\caption{Thermal bolometric light curves of thermal radiation from a NS-NS merger leading to a long-lived NS. A possible contribution from the spin-down energy is included.}
\label{NSbol}
\end{center}
\end{figure}

In Figure~\ref{NSbol}, we show thermal bolometric light curves of the thermal luminosity, $L_{\rm th}$.  The ejecta mass and velocity are assumed to be $M=0.02~M_\odot$ and $V=0.2~c$, respectively. If a long-lived NS exists with $P_i\sim1$~ms, the energy injection from the pulsar can readily exceed that from radioactive nuclei unless the magnetic field is less than $\sim10^{10}$~G. The spin-down time can be much shorter than the photon breakout time if the magnetic field is strong enough, in which case the radiation luminosity is not far from that of the radioactive-decay-powered one. However, the ejecta speed becomes close to $c$, which is at least different from the case of GW+EM170817. The case of $B_*=5\times{10}^{16}$~G and $P_i=10$~ms, which is motivated by SGRB extended emission, keeps the ejecta speed comparable to the original ejecta velocity and could put an energy injection at early times. 
Even if the long-lived pulsar model does not explain GW+EM170817, such a long-lived NS remnant could be born in low-mass NS-NS binaries. Our study here allows for the possible variety of NS remnants. 

\begin{figure}[tb]
\begin{center}
\includegraphics[width=\linewidth]{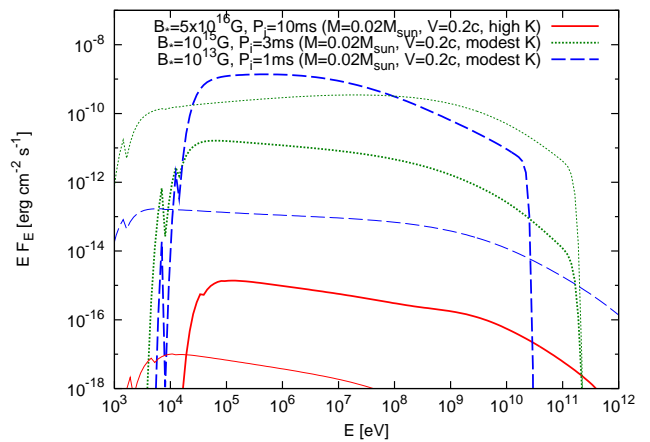}
\caption{High-energy photon spectra of a long-lived pulsar left as a compact remnant of the NS-NS merger, at $t=10^6$~s (thick curves) and $t=10^7$~s (thin curves). The distance is set to $d=40$~Mpc.}
\label{NSspec}
\end{center}
\end{figure}

In Figure~\ref{NSspec}, we show spectra of embryonic pulsar wind nebulae embedded in the merger ejecta. The spectra consist of synchrotron and inverse-Compton components, and the latter is dominant in the GeV-TeV range~\citep{MKK15a}. In the case of $B_*={10}^{15}$~G and $P_i=3$~ms, the external inverse-Compton component due to up-scattering of thermal photons is prominent, making the broadband spectrum as flat as $EF_E\propto{\rm const.}$ even beyond $E\sim m_ec^2$. Hard X-ray emission with a very hard spectrum is also expected above $\sim10-100$~keV. Note that the rotation energy for $P_i\lesssim10$~ms is ${\mathcal E}_{\rm rot}\gtrsim4\times{10}^{51}$~erg, which can exceed $\sim5{\mathcal E}_{\rm ej}=(5/2)MV^2\sim4\times{10}^{51}$~erg~\citep{Suzuki:2016gbg}, so a wind bubble breakout occurs through Rayleigh-Taylor instabilities, which allows more X-rays to escape, would occur. Here ${\mathcal E}_{\rm ej}$ is the ejecta kinetic energy. 
Note that the gamma-ray attenuation by the extragalactic background light is not included in this work. 

\begin{figure}[tb]
\begin{center}
\includegraphics[width=\linewidth]{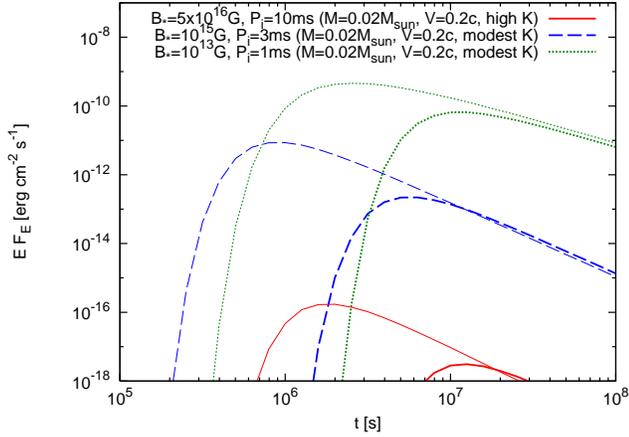}
\caption{X-ray light curves from a long-lived pulsar as a merger remnant, for $E=3$~keV (thick curves) and $E=30$~keV (thin curves). }
\label{NSX}
\end{center}
\end{figure}

\begin{figure}[tb]
\begin{center}
\includegraphics[width=\linewidth]{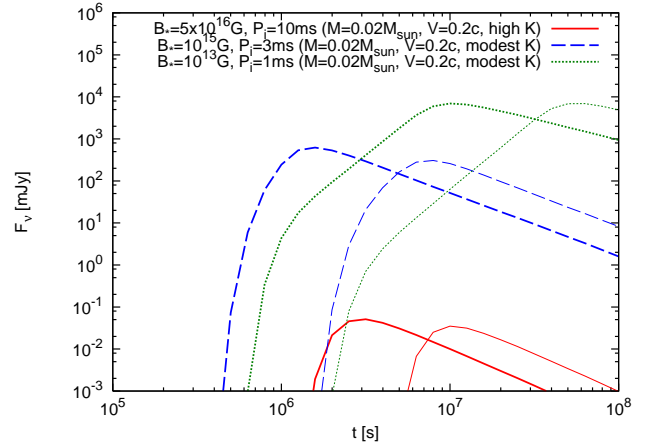}
\caption{High-frequency radio light curves from a long-lived pulsar as a merger remnant, for $\nu=100$~GHz (thick curves) and $\nu=10$~GHz (thin curves).}
\label{NSradio}
\end{center}
\end{figure}

We show the X-ray light curves in Figure~\ref{NSX} and the radio light curves in Figure~\ref{NSradio}. Using {\it NuSTAR}, with sensitivity $EF_E\sim{10}^{-14}~{\rm erg}~{\rm cm}^{-2}~{\rm s}^{-1}$, hard X-ray emission from the embryonic nebula with $B_*={10}^{15}$~G and $P_i=3$~ms is detectable up to $z\sim0.2-0.3$. The detection prospects are quite sensitive to the spin-down parameters. 
We find that radio and sub-mm observations are more promising, which is consistent with \cite{Murase:2016sqo}, who proposed synchrotron nebular emission as a probe of the connection between fast radio bursts and pulsar-driven supernovae, including super-luminous supernovae. 
High-frequency radio emission can escape around $\sim{10}^{6}-{10}^{7}$~s thanks to the small ejecta mass, the fast velocity, and the expectation that the ejecta are largely neutral. In the case where $B_*={10}^{15}$~G and $P_i=3$~ms, the sub-mm emission can be detected up to $z\sim1.5$ by ALMA with sensitivity of $\sim0.01$~mJy.  
Note that the radio synchrotron spectrum (which cannot be harder than $F_\nu\propto\nu^{-0.5}$) is $F_\nu\sim\nu^{-0.8}-\nu^{-0.7}$ in our cases~\citep[see][for a detailed discussion]{Murase:2016sqo}, and it declines as $F_\nu\propto t^{-2}$. 
The long-lived pulsar model can be discriminated from the BH disk wind model, the merger ejecta shock model, and the GRB afterglow model, by using the spectral index and the time evolution. 

\begin{figure}[tb]
\begin{center}
\includegraphics[width=\linewidth]{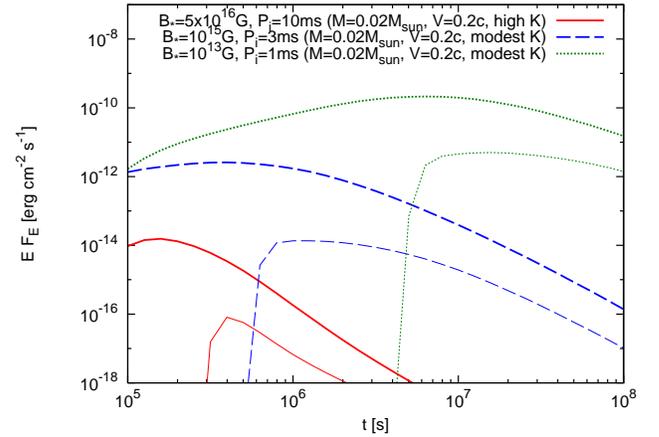}
\caption{Gamma-ray light curves from a long-lived pulsar as a merger remnant, for $E=1$~GeV (thick curves) and $E=100$~GeV (thin curves).}
\label{NSgamma}
\end{center}
\end{figure}

Finally, we show gamma-ray light curves in Figure~\ref{NSgamma}. The gamma-ray breakout time obtained by numerical calculations is consistent with the analytical estimate given in Equation~(\ref{gammatime}). 
For $P_i\sim1-3$~ms and $B_*\sim{10}^{13}-{10}^{15}$~G, the GeV gamma-ray flux is estimated to be $EF_E\sim{10}^{-12}-{10}^{-10}~{\rm erg}~{\rm cm}^{-2}~{\rm s}^{-1}~{(d/40~{\rm Mpc})}^{-2}$, which can be detected by {\it Fermi} which has sensitivity $EF_E\sim{10}^{-11}~{\rm erg}~{\rm cm}^{-2}~{\rm s}^{-1}$ in the GeV range. 
TeV emission is usually suppressed by the Klein-Nishina effect, which makes detections more challenging. But such nebular emission can be much brighter than the forward shock emission by the merger ejecta~\citep{Takami:2013rza}.   
More generally, we conclude that gamma-ray detection of a pulsar remnant is possible when the spin-down time is sufficiently long, in which case very bright optical transients will also be present (cf. Figure~\ref{NSbol}). 

\subsection{Implications from X-Ray and Radio Observations of GW+EM170817}
In the previous sections, we have studied non-thermal emission expected in the post-merger phase. While our purpose is to provide a general study rather than a specific study on GW+EM170817, it would be interesting to discuss the consequences for this object.

X-ray observations have been reported by various authors~\citep{Evans:2017aa,Margutti:2017aa,Troja:2017nqp}. In particular, {\it Chandra} detected weak X-ray signals with $EF_E\sim5\times{10}^{-15}~{\rm erg}~{\rm cm}^{-2}~{\rm s}^{-1}$, 9~d and 15~d after the GW and GRB events. The non-detection of earlier X-ray emission excludes on-axis afterglow emission from a highly relativistic jet. While the observed emission could be explained by mildly relativistic outflows~\citep{Evans:2017aa,Troja:2017nqp}, the most natural explanation for these observations is off-axis afterglow emission by a canonical SGRB jet, with a viewing angle of $\theta\sim30$~deg~\citep{Kim:2017skw,Margutti:2017aa,Troja:2017nqp}. Intriguingly, the same afterglow model with an isotropic-equivalent kinetic energy of ${\mathcal E}_k\sim{10}^{50}~{\rm erg}$ and an ambient density of $n\sim{\rm a~few}\times{10}^{-3}~{\rm cm}^{-3}$ also explains the radio signals, which were detected with $F_\nu\sim0.03-0.04~{\rm mJy}$ at 3~GHz and 6~GHz, $\sim{10}^6$~s after the merger~\citep{Alexander:2017aly,Hallinan:2017aa}. 

These X-ray and radio data as well as non-detection by other facilities at different epochs enable us to place interesting upper limits on long-lasting activity of the central engine. {\it Swift} and {\it NuSTAR} have provided upper limits with $EF_E\lesssim(3-5)\times{10}^{-14}~{\rm erg}~{\rm cm}^{-2}~{\rm s}^{-1}$, from $\sim1$~d to $\sim30$~d after the coalescence~\citep{Evans:2017aa}, and late-time observations are especially important for the purpose of searching for the compact merger remnant (e.g., the upper limit by {\it NuSTAR} at $t\sim30$~d). 
These upper limits can now be compared to the results shown in Figure~\ref{BHX}. The disk emission model predicts that X-ray emission can escape $\sim30-100$~d after the merger, so the X-ray emission observed by {\it Chandra} is unlikely to be the remnant origin. {\it NuSTAR} provided an interesting upper limit, $EF_E\lesssim3\times{10}^{-14}~{\rm erg}~{\rm cm}^{-2}~{\rm s}^{-1}$, at $t\sim30$~d. While this is consistent with our fiducial case, the {\it NuSTAR} upper limit implies that the disk luminosity should be constrained to be $L_{\rm disk}\lesssim10^{41}~{\rm erg}~{\rm s}^{-1}$. Thus, based on Equation~(\ref{Ldiskmax}), the fall-back disk with $L_{\rm disk}\gtrsim100L_{\rm Edd}$ is ruled out. 
This leads to constraint, $\eta_{\rm rad}\lesssim100$, if $t<t_{\rm tr}$, i.e., $\dot{M}_0\gtrsim3\times{10}^{-3}~M_\odot~{\rm s}^{-1}$. Or we simply have $\dot{M}_0\lesssim3\times{10}^{-3}~M_\odot~{\rm s}^{-1}$.
Interestingly, Figure~\ref{BHX} suggests that further late-time observations can be critical. The X-ray emission from the fall-back disk in the keV range should be suppressed at relatively early times, but it would become prominent at $t\sim{10}^7~{\rm s}\sim100~{\rm d}$. With a deep observation by {\it Chandra}, with an energy sensitivity of $EF_E\sim{10}^{-15}~{\rm erg}~{\rm cm}^{-2}~{\rm s}^{-1}$ (in an integration time of $3\times{10}^5$~s), the associated X-ray signal may be detected; otherwise the non-detection will give us an upper limit of $L_{\rm disk}\lesssim10L_{\rm Edd}$, leading to $\eta_{\rm rad}\lesssim10$ or $\dot{M}_0\lesssim3\times{10}^{-2}~M_\odot~{\rm s}^{-1}$.

We also point out that late-time optical/IR observations give us independent constraints. Since X-ray emission is absorbed in the ejecta at early times, the disk emission model cannot have a high luminosity. From Figure~\ref{BHlc}, we find that the observed bolometric luminosity in the optical/IR band at $\sim10-20$~day suggests $L_{\rm disk}\lesssim(10-20)~L_{\rm Edd}$~\citep[e.g.,][]{Kilpatrick:2017mhz}. This also indicates that further late-time observations in the IR band give us useful information. However, since X-rays start to escape at $t_{HX-\rm thin}\sim10-50$~d, detailed studies on the thermalization in the ejecta are necessary to place reliable constraints.   

X-ray observations can also constrain the pulsar model. While such a model may be disfavored in light of neutrino effects on the $r$-process nucleosynthesis~\citep{Lippuner:2017bfm}, X-ray observations independently require that the energy injection time ($t_{\rm inj}$) be sufficiently long, and/or the injection luminosity ($L_{\rm inj}$) be sufficiently low. If $t_{\rm inj}\approx t_{\rm collapse}$, Figure~\ref{NSX} suggests that the spin-down parameters motivated by SGRB extended emission can easily avoid such constraints.   

High-frequency radio observations may be more powerful for testing the long-lived pulsar model. Figure~\ref{NSradio} indicates that bright synchrotron nebular emission can be expected in this model. In particular, ALMA reported interesting upper limits, $F_\nu\lesssim0.1$~mJy at 338.5~GHz, $t\sim4\times{10}^6$~s after the coalescence~\citep{Kim:2017skw}, and $F_\nu\lesssim0.04$~mJy at 97.5~GHz, $t\sim15-30$~d after the coalescence~~\citep{Alexander:2017aly}. 
The cases with $(B_*,P_i)=({10}^{15}~{\rm G},3~{\rm ms})$ and $(B_*,P_i)=({10}^{13}~{\rm G},1~{\rm ms})$ are clearly ruled out, as already indicated by Figure~\ref{NSbol} via optical/IR observations. Interestingly, we find that even the case of $B_*=5\times{10}^{16}$~G and $P_i=10$~ms (where the parameters are motivated by SGRB extended emission while the bolometric luminosity remains consistent with the observations) is only marginally consistent with the ALMA upper limit. More detailed parameter scans are beyond the scope of this work, but this demonstrates that the existence of the long-lived pulsar can be constrained independently of the GW signals and the optical/IR emission (that originates from the thermalization of the nebular emission).    

High-frequency radio emission can be expected even if the remnant is a BH in the dissipative outflow model. As shown in Figure~\ref{BHradio}, the synchrotron flux from the BH wind nebula is $F_\nu\sim0.001$~mJy, which is consistent with the current upper limits although further late-time radio observations could be more useful.  Note that in this model, the outflow rate should satisfy $\eta_w\dot{M}_d\lesssim3\times{10}^{-3}~M_\odot~{\rm s}^{-1}$ for $V_w=0.3~c$, otherwise the thermal bolometric luminosity would be affected.  

Note that radio signals from wind nebulae should be quite different from those produced by the off-axis jet~\citep[e.g.,][]{vanEerten:2010zh} and merger ejecta~\citep[e.g.,][]{NP11a}, in the sense that low-frequency radio emission should be suppressed more strongly. 
Indeed, the spectral index of the radio signal at 3 and 6 GHz does not show absorption features~\citep{Hallinan:2017aa} so that the early radio emission arises from the optically-thin forward shock of the SGRB afterglow.

\section{GeV-TeV Gamma-Rays?}
The high-energy signatures discussed in the previous section are expected even for off-axis observers. Whereas the X-ray and radio signals can be used as a powerful probe, our results suggest that GeV-TeV gamma-rays coming directly from the hidden merger remnant are difficult to detect, at least for the parameters favored by GW+EM170817. 
In this section, we discuss high-energy emission that involves a SGRB jet. Such emission is the most powerful for on-axis observers, but weaker signals may be detected for off-axis observers.     

\subsection{Up-scattered Long-Lasting Engine Emission}
For an on-axis observer, extended and plateau emission from SGRBs have been observed as well as X-ray flares, so these emission components are most naturally explained as late-time internal dissipation via internal shocks or magnetic reconnections or photospheric dissipation~\citep[e.g.,][]{GGN07a,MTY11a}. 
It seems unlikely that every merger event has an isotropic luminosity comparable to that of the observed luminosity of late-prompt emission from SGRBs~\citep[][]{Fong2014}. At least, both X-ray and optical/IR observations of GW+EM170817 strongly constrain underlying X-ray components, indicating that long-lasting SGRBs emission should be beamed if it exists. 

If the dissipation region is close to the central engine or is highly relativistic, off-axis observers cannot observe such long-lasting X-ray and gamma-ray emission. 
On the other hand, a relativistic jet that is responsible for prompt SGRB emission is decelerated so quickly that its Lorentz factor would become relatively low. Indeed, the deceleration time for an {\it on-axis observer} is estimated to be $T_{\rm dec}\simeq10~{\rm s}~{\mathcal E}_{k,50}^{1/3}n_{-3}^{-1/3}{(\Gamma_0/300)}^{-8/3}$, where ${\mathcal E}_k$ is the isotropic-equivalent kinetic energy, $n$ is the ambient density, and $\Gamma_0$ is the initial bulk Lorentz factor. For an adiabatic relativistic blast wave expanding into environments with constant density, the bulk Lorentz factor is estimated to be $\Gamma (T) \simeq 44~\mathcal{E}_{k,50}^{1/8} n_{-3}^{-1/8} T_{3}^{-3/8}$, and the corresponding external shock radius is given by $R(T) \simeq 2.3 \times {10}^{17}~{\rm cm}~ \mathcal{E}_{k,50}^{1/4} n_{-3}^{-1/4} T_{3}^{1/4}$. 

The standard afterglow model~\citep[for a review see][and references therein]{Meszaros:2006rc} has successfully explained multi-wavelength data of GRBs including SGRBs. Non-thermal electrons are accelerated with an injection Lorentz factor of $\gamma_{e,i}\simeq2.3\times{10}^{3}\epsilon_{e,-1}f_{e}^{-1}(g_s/g_{2.4})\mathcal{E}_{k,50}^{1/8}n_{-3}^{-1/8}T_3^{-3/8}$, where $f_e$ is the number fraction of non-thermal electrons and $g_s=(s-1)/(s-2)$.
The cooling Lorentz factor of electrons is estimated to be $\gamma_{e,c}\simeq6.2\times{10}^{3}\epsilon_{B,-2}^{-1}\mathcal{E}_{k,50}^{-3/8}n_{-3}^{-5/8}T_3^{1/8}{(1+Y)}^{-1}$~\citep[e.g.,][]{2002ApJ...568..820G}, where $Y$ is the total Compton $Y$ parameter. 

High-energy gamma-rays can be produced by the inverse-Compton radiation process, in particular in afterglow shocks with external photons from the long-lasting jet~\citep[e.g.,][]{Meszaros+93multi,Fan+2008a}. In the presence of late prompt emission (i.e. extended emission, plateau emission, and X-ray flares), not only synchrotron self-Compton but also external inverse-Compton processes by forward external shock electrons become important. The extended emission and plateau components can be described by a broken power law, 
\begin{equation}
EF_E^{\rm LP}(T)\propto\frac{L_{\rm LP}}{4\pi d^2}\propto
\left\{ \begin{array}{ll} 
T^{-\alpha_{\rm fl}}
& \mbox{($T < T_{a}$)}\\
T^{-\alpha_{\rm st}}
& \mbox{($T_a \leq T$)},
\end{array} \right.
\end{equation}
where $EF_E^{\rm LP}$ is the energy flux, $L_{\rm LP}$ is the luminosity, $T_a \sim {10}^{2.5}$~s is the break time for extended emission and $T_a\sim{10}^{4}$~s is for plateau emission. Throughout this work, we use $\alpha_{\rm fl}=0$ and $\alpha_{\rm st}=40/9$~\citep{KI15a,KIS17a}. We use a luminosity at the break energy, $L_{\rm LP}^b={10}^{49}~{\rm erg}~{\rm s}^{-1}$, and $E^b=1$~keV for the extended emission, while we adopt $L_{\rm LP}^b={10}^{47}~{\rm erg}~{\rm s}^{-1}$ and $E^b=0.1$~keV for the plateau emission. 

For a given time-dependent seed photon field from an inner source close to the central engine, the external inverse-Compton emission spectrum observed for an on-axis observer is calculated by the following formula~\citep{MTY11a,Murase:2009su},
\begin{eqnarray}
F_E^{\rm EIC} (T)&=&\frac{3}{2} \sigma_T \int \frac{d r}{r} (1-\cos \tilde{\vartheta}) \int d \gamma_e \, \frac{d n_e}{d \gamma_e} \tilde{\Delta} \int d y \, (1-\xi) \nonumber \\ 
&\times& \left[ 1- 2 y +2 y^2 + \frac{\xi^2}{2(1-\xi)} \right]
\frac{F_{E}^{{\rm LP},b} (r) G(\varepsilon)}{{(1+\Gamma^2 \vartheta^2)}^2}
\,\,\,\,\,
\label{EICformula}
\end{eqnarray}
where 
$y \equiv \frac{\xi m_e c^2}{2(1-\cos \tilde{\vartheta}) \gamma_e \varepsilon (1-\xi)}$ and $\xi \equiv \frac{(1+ \Gamma^2 \vartheta^2) E}{2 \Gamma \gamma_e m_e c^2}$. The distribution of electrons in the comoving frame is $dn_e/d \gamma_e \propto \gamma_e^{-s}$, and $\tilde{\Delta}\approx\Gamma cT$ is the comoving shell size. The scattering angles $\vartheta$ and $\tilde{\vartheta}$ of external inverse-Compton photons are measured in the central engine frame and the comoving frame, respectively. 
The function $G(\varepsilon)$ represents the spectrum of seed photons with energy $\varepsilon$ in the comoving frame. We use $G(\varepsilon)={(\varepsilon/\varepsilon^b)}^{-\beta_l+1}$ for $\varepsilon < \varepsilon^b$ and $G(\varepsilon)={(\varepsilon/\varepsilon^b)}^{-\beta_h+1}$ for $\varepsilon^b \leq \varepsilon$, respectively, with $\beta_l=0.5$ and $\beta_h=2.0$ as photon indices. 

For viewing angle, $\theta$, and jet opening angle, $\theta_j$, we calculate emission for off-axis observers in a simplified approach~\citep{Granot:2002za}. Introducing $\eta\equiv\Gamma(1-(v/c))\delta(\Delta\theta)$ (where $v$ is the jet velocity corresponding to $\Gamma$), we use the relationship, $F_{E}(t)={\eta}^{3}F_{E/\eta}(\eta t)$, where $\Delta \theta\equiv\theta-\theta_j$ and $\delta(\Delta\theta)={[\Gamma(1-(v/c)\cos\Delta\theta)]}^{-1}$~\citep[see also][]{2001ApJ...554L.163I}. Although this prescription for a top-hat jet only gives approximate results, it is sufficient for the purpose of this work. For more detailed evalulation, one needs to calculate the equal-arrival-time surface for an off-axis observer, taking into account the anisotropy of inverse-Compton scattering, as well as the possible jet structure~\citep[e.g.,][]{Meszaros:1997je}.  

\begin{figure}[tb]
\begin{center}
\includegraphics[width=\linewidth]{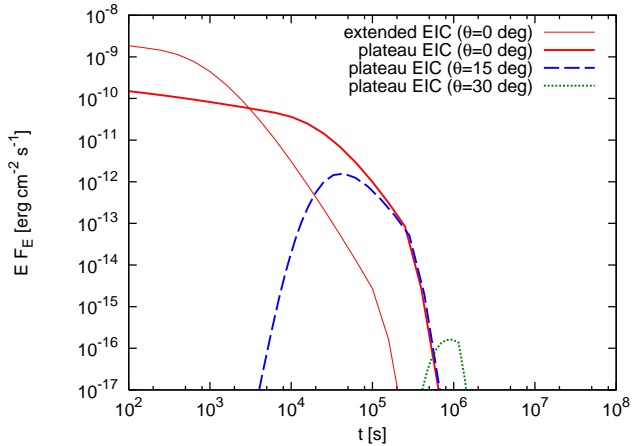}
\caption{Light curves of high-energy gamma-rays generated by external inverse-Compton radiation, for $E=100$~GeV. Three different viewing angles (measured from the jet axis) are considered, and extended emission with $T_a={10}^{2.5}$~s and plateau emission with $T_a={10}^{4}$~s are assumed as seed photons. The distance is set to $d=40$~Mpc.} 
\label{EIClc}
\end{center}
\end{figure}

\begin{figure}[tb]
\begin{center}
\includegraphics[width=\linewidth]{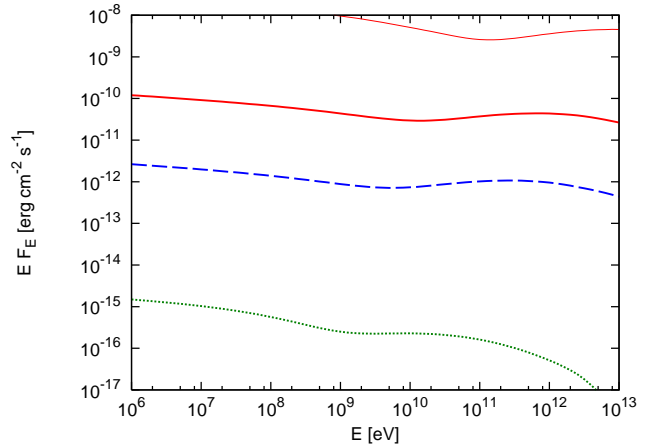}
\caption{Gamma-ray spectra corresponding to Figure~\ref{EIClc}. 
For extended emission with $T_a={10}^{2.5}$~s, the spectrum at $t=10^2$~s is shown (top curve), and the viewing angle is set to $\theta=0$~deg.
For plateau emission with $T_a={10}^{4}$~s, the spectra at $t=10^4$~s, $t=2.5\times10^4$~s, and $t=8.2\times10^5$~s are  shown (from the second top to bottom), and the viewing angles are $\theta=0$~deg, $\theta=15$~deg, and $\theta=30$~deg, respectively.}
\label{EICspec}
\end{center}
\end{figure}

The results are given in Figures~\ref{EIClc} and \ref{EICspec}. The afterglow parameters are ${\mathcal E}_k=2\times{10}^{50}$~erg, $n=10^{-3}~{\rm cm}^{-3}$, $\epsilon_e=0.1$, $\epsilon_B=0.01$, $s=2.2$, and $\theta_j=0.2$. For an on-axis observer, the duration of the external inverse-Compton radiation is comparable to $T_a$. 
The resulting inverse-Compton radiation is so bright that it is detectable up to $\sim300$~Mpc with {\it Fermi}-LAT, HAWC, and current imaging atmospheric Cherenkov telescopes such as MAGIC, VERITAS, HESS, and CTA in future. In particular, CTA is expected to be powerful due to its high sensitivity of $EF_E\sim{10}^{-12}~{\rm erg}~{\rm cm}^{-2}~{\rm s}^{-1}$ at 100~GeV and $EF_E\sim{10}^{-13}~{\rm erg}~{\rm cm}^{-2}~{\rm s}^{-1}$ at 1~TeV~\citep{2013APh....43..252I}. 
On the other hand, the duration of off-axis emission is significantly longer, of order $t\sim1-10$~d for $\theta\sim15-30$~deg. However, in this simple top-hat jet model, the expected gamma-ray flux becomes significantly lower as the viewing angle is larger than $\theta_j$. 

\subsection{Implications from GeV-TeV gamma-ray observations of GW+EM170817}
An MeV gamma-ray counterpart of GW+EM170817 was detected, which is identified with GRB 170817A~\citep{GBM:2017lvd,Savchenko:2017ffs}. The origin of the MeV gamma-ray emission, which is under debate, could be, e.g., off-axis jet emission, or on-axis emission from mildly relativistic outflows including a cocoon formed by a jet drilling through the merger ejecta. On the other hand, GeV gamma-rays associated with GRB 170817A have not been detected by observations, and there are only upper limits, e.g., $EF_E\lesssim2\times{10}^{-9}~{\rm erg}~{\rm cm}^{-2}~{\rm s}^{-1}$ in the $1-100$~GeV range for $t_{\rm GW}+1153$~s to $t_{\rm GW}+2027$~s~\citep{Fermi-LAT:2017uvi}.   

For the purpose of searching for long-lasting gamma-ray emission described in this paper, we have also analyzed the {\it Fermi}-LAT data positionally coincident with the optical counterpart of the LIGO event (${\rm RA}=197.45^{\circ}, {\rm DEC}=-23.3815^{\circ}$) and temporally selected between 2017 August 17 to 2017 September 17 (524620805 to 527299205 MET). The analysis was performed with \textit{ScienceTools v10r0p5} using the \textit{P8R2\_SOURCE\_V6} instrument response function. Events were selected from within a $21.2^{\circ}\times21.2^{\circ}$ region, centered on the counterpart position, defining our region of interest (ROI). Further cuts were made by selecting \textit{SOURCE} class photons of energies ranging from 0.1 -- 300 GeV and filtered for data-taking periods corresponding to good time intervals using \textit{gtmktime}. Data were then binned spatially in $0.1^{\circ}$ sized-pixels and in energy with 34 logarithmically uniform bins. A $50^{\circ}\times50^{\circ}$ exposure map was created, centered on the source position, using the same binning as the ROI. A larger exposure map was used to account for potential contributions from sources not in our ROI -- a consequence of {\it Fermi}-LAT\textsc{\char13}s large point-source spreading function at low energies.

\begin{figure}[tb]
\begin{center}
\includegraphics[width=\linewidth]{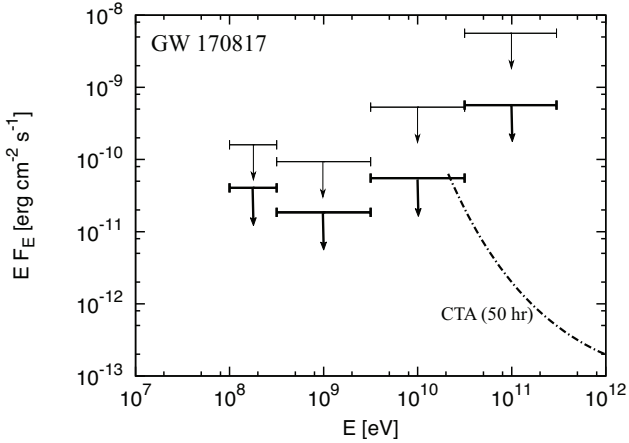}
\caption{The gamma-ray upper limits by the {\it Fermi}-LAT data, obtained for GW+EM170817. The data were analyzed for 1~d (thin data) and 10~d (thick data) since August 17 in 2017. The CTA sensitivity is overlaid for comparison.}
\label{fermi}
\end{center}
\end{figure}

The region is modelled with all known 3FGL sources within $25^{\circ}$ along with the Galactic, \textit{gll\_iem\_v06}, and isotropic, \textit{iso\_P8R2\_SOURCE\_V6\_v06}, diffuse emission templates. Except for the normalization of the variable 3FGL sources and the isotropic diffuse emission, all other source parameters were fixed. The normalization of the Galactic diffuse model was fixed given the source\textsc{\char13}s high Galactic latitude ($b = 39.3^{\circ}$). The event is modeled with a simple power law where the normalization was left free while the photon index and pivot energy were fixed to $\gamma = 2.0$ and 100 MeV, respectively. The integrated flux and significance were then determined using a binned likelihood approach with the \textit{pyLikelihood} module and \textit{BinnedAnalysis} functions, utilizing the \textit{NewMinuit} optimizer. We correct for energy dispersion, given LAT’s poor energy resolution for $E\lesssim 300$ MeV photons, during the likelihood fitting. Our fits show no significant gamma-ray emission from such a source at this position. 
A $95\%$ upper limit was calculated for the source using \textit{IntegralUpperLimit}; a Python module part of the Fermi \textit{ScienceTools} package. We calculated a time integrated spectrum for 1 and 10 days after the event, with the same procedure as above, for 4 energy bins (see Figure~\ref{fermi}). Note that our results are consistent with the independent analyses by the {\it Fermi}-LAT Collaboration~\citep{Fermi-LAT:2017uvi}. 

The gamma-ray limits presented in Figure~\ref{fermi} are consistent with the off-axis predictions of external Compton emission shown in Figures~\ref{EIClc} and \ref{EICspec}. HESS reported an upper limit on the TeV gamma-ray flux, $EF_E\lesssim{10}^{-12}~{\rm erg}~{\rm cm}^{-2}~{\rm s}^{-1}$~\citep{HESS:2017aa}, which is also consistent with the off-axis interpretation of the canonical jet with $\theta\gtrsim15$~deg. 
Gamma-ray emission from a long-lived pulsar, which is shown in Figure~\ref{NSgamma}, is also consistent with the non-detections of GeV-TeV gamma-rays. 

\section{Summary and Discussion}
%
The recent observations of GW+EM170817 support the physical link between SGRBs and NS-NS mergers, but details of their connection are not yet understood. To reveal the compact remnants from NS-NS mergers and figure out their roles as the central engine of ``long-lasting'' SGRB emission, we investigated the high-energy signatures originating from a BH with a disk and a long-lived pulsar. We showed that GeV-TeV gamma-rays are produced as a consequence of such prolonged activity of the central engine at the external shocks, through up-scatterings of late-prompt photons generated by internal dissipation in the jet.  

Regarding the late-time emission from the BH-disk system, we considered two possibilities, the disk emission model and the dissipative outflow model. In the former model, if the accretion rate is as high as the value required to explain the ubiquitous SGRB plateau emission and late-time flares, X-rays from the disk or the corona can be seen for nearby merger events at $30-100$~d after the coalescence. Deep X-ray observations reaching $L_{\rm disk}\sim L_{\rm Edd}$ are desirable, with which we can examine whether the accretion mode is super-Eddington or not, and which will enable us to constrain $\dot{M}_d$ and to test the remnant disk model as the central engine of long-lasting SGRB emission. 
In future, more NS-NS merger events will be detected by LIGO and Virgo, and ultimate X-ray missions such as {\it Athena+} will enable us to test the late-time accretion scenario for long-lasting SGRB emissions and probe the fall-back process in NS-NS mergers.   
The X-ray and optical/IR observations of GW+EM170817 indicate $L_{\rm disk}\lesssim(10-100)L_{\rm Edd}$, and we encourage further late-time observations by {\it Chandra} and by IR telescopes to test the possibility of the long-lasting accretion onto the central remnant. With the data that are currently available, we cannot yet exclude the possibility that super-Eddington accretion occurs. But, even so, $\eta_{\rm rad}\gtrsim10-100$ is unlikely. The observations are also consistent with the theoretically inferred fall-back rate. On the other hand, disk-driven outflows may exceed the super-Eddington rate, but the compatibility with the kilonova/macronova emission suggests $\dot{M}_0\lesssim{10}^{-2}~M_{\odot}~{\rm s}^{-1}$, if fall-back evolution is assumed. In the latter model, high-frequency radio signals from the BH wind nebula can be expected as well as ordinary radio emission from the forward shock of the merger. 

As an alternative model, we also calculated non-thermal nebular emission from the long-lived pulsar system. Given that the pulsar is an efficient accelerator of electrons and positrons, as suggested by the observations of Galactic pulsar wind nebulae, not only X-rays but also gamma-rays and radio waves from embryonic pulsar wind nebulae may serve as signatures of the long-lasting central engine. Although the expected fluxes are sensitive to the spin-down parameters, we found that strongly magnetized pulsars with $P_i\sim1-3$~ms lead to bright nebular emission even in the high-frequency radio and in the gamma-ray range. 
However, there have been no observational indications of such embryonic nebulae, for both SGRBs and NS-NS mergers such as GW+EM170817. Even in the case of GW+EM170817, we demonstrated that high-frequency radio observations start to constrain the pulsar-driven model for the spin-down parameters motivated by SGRB extended emission. This is complementary to the previous efforts in the context of SGRBs, in which radio observations have provided meaningful constraints on the possibility that the ejecta are mainly powered by the spin-down energy~\citep{Metzger:2013cka,Horesh:2016dah}. These implications of GW+EM170817 could imply that the massive NS may have collapsed into a BH after a while or the large-scale magnetic field is extremely weak.

So far, we focused on the BH disk model and long-lived pulsar model, both of which has been suggested as a possible explanation for SGRB long-lasting emission. On the other hand, the former model requires a higher accretion rate while the origin of the large rotation energy is unclear in the latter model. 
These could be overcome by a hybrid model involving both a spinning NS and BH with a remnant disk. The massive NS may act as the magnetar central engine to produce relativistic jets~\citep[e.g.,][]{Usov:1992zd}, and eventually become a BH after losses of its angular momentum. Phenomenologically, the injection luminosity and the energy injection time are written as
\begin{equation}
L_{\rm inj}\approx L_{\rm sd}
\end{equation}
and
\begin{equation}
t_{\rm inj}\approx t_{\rm collapse},
\end{equation} 
where $t_{\rm collapse}$ is the BH formation time and only a fraction of $L_{\rm sd}$ may be injected into the quasi-isotropic ejecta. Extended emission could be attributed to $t_{\rm collapse}$, during which the fall-back accretion may be prevented by the proto-NS wind or heating from the massive NS. Other late-time emission such as plateaus or flares may be attributed to the delayed accretion onto the BH. Scanning the phenomenological parameter space of $L_{\rm inj}$ and $t_{\rm inj}$ are left for future work. Note that the above hybrid scenario can also be tested by late-time X-ray and optical/IR observations, as discussed in Section~2.1.  

If $t_{\rm collapse}$ is shorter than any of the diffusion time for thermal photons, $t_{HX-\rm thin}$, $t_{\gamma-\rm thin}$ and so on, then high-energy signatures of massive NSs would be more difficult to detect by electromagnetic wave observations (although dedicated observations of GWs are promising). 
Nevertheless, there may be diversity in the NS-NS binaries, in which case we would expect different kinds of transients. In particular, a low-mass NS-NS binary could end with the coalescence leaving a long-lived NS, which could be accompanied by bright, multi-wavelength nebular emission .
Interestingly, the properties of non-thermal nebular emission are quite similar to those studied in the context of fast radio bursts and super-luminous supernovae~\citep[][]{Murase:2016sqo}, except that the ejecta mass is larger and the composition is heavier. If young neutron stars (including magnetars and fast-spinning pulsars) are responsible for fast radio bursts, the remnants of the low-mass NS-NS binaries could be the sources of fast radio bursts, and association with fast-cooling synchrotron nebular emission, which we studied in this work, is expected. Pulsar-driven optical transients have been expected not only at the end of the NS-NS system~\citep[e.g.,][]{Yu:2013kra,Metzger:2014ks,Gao:2015rua,DOrazio:2015jcb} but also at the birth of the NS-NS system~\citep[e.g.,][]{Hotokezaka:2017qng}.

Non-thermal emission at different energies will be useful for probing the properties of the merger ejecta. Optical/IR photons, X-rays, and radio waves interact with the ejecta material differently, depending on composition and ionization. Gamma-rays can propagate without dependence on the details of the atomic states, which potentially enables us to probe deeper regions closer to the central engine. Neutrinos have much higher penetration power, which may enable us to get more critical information on the central engine~\citep[e.g.,][]{Murase:2009pg,Fang:2017tla}.

\begin{acknowledgements}
We thank Teruaki Enoto for useful discussions. 
The work of K.M. is supported by Alfred P. Sloan Foundation and NSF grant No. PHY-1620777. S.S.K. acknowledges the IGC fellowship at Penn State University.
This work is partly supported by ``New Developments in Astrophysics Through Multi-Messenger Observations of Gravitational Wave Sources'', No. 24103006 (K.I.), KAKENHI Nos. 26287051, 26247042, 17H01126, 17H06131, 17H06362, 17H06357 (K.I.), and 17K14248 (K.K.), by the Grant-in-Aid from the Ministry of Education, Culture, Sports, Science and Technology (MEXT) of Japan, and by NASA NNX13AH50G (P.M.).
We thank the YITP workshop (YITP-W-16-15) on electromagnetic counterparts of gravitational wave sources (held in March 2017), where this project was initiated.
\end{acknowledgements}

\bibliographystyle{apj_8}
\bibliography{merger}

\end{document}